\documentclass[aip, rsi, amsmath,amssymb, reprint,]{revtex4-1}

\usepackage{graphicx}% Include figure files
\usepackage{dcolumn}% Align table columns on decimal point
\usepackage{bm}% bold math
\usepackage{caption}
\usepackage{subcaption}
\usepackage[sort&compress]{natbib}
\usepackage{afterpage}
\usepackage{epstopdf}
\usepackage{etoolbox}

\def \figuresize {8cm}
\def \HW {half waveplate}
\def \MI {Montana Instruments}
\def \WN {cm$^{-1}$}
\def \FR {Fresnel Rhomb}
\newcommand{\BS}{ Bi$_{2}$Se$_{3}$}
\newcommand{\VO}{V$_{2}$O$_{3}$}

\newcommand{\comment}[2]{#2}
\begin{document}

\preprint{AIP/123-QED}

\title[Low Vibration high NA automated variable temperature Raman microscope]{Low Vibration high NA automated variable temperature Raman microscope}
%\thanks{Footnote to title of article.}

\author{Yao Tian}
\thanks{These two authors contributed equally}
%\thanks{These three authors contributed equally}
 \affiliation{Department of Physics \& Institute for Optical Sciences, University of Toronto, 60 St. George Street, Toronto, ON M5S 1A7, Canada}

\author{Anjan A. Reijnders}
\thanks{These two authors contributed equally}
 \affiliation{Montana Instruments, 151 Evergreen Dr., Bozeman, Montana 59715 USA.}

%\author{Michael E. Holmes}
 %\affiliation{Montana Instruments, 151 Evergreen Dr., Bozeman, Montana, USA.}

\author{Gavin B. Osterhoudt}
\affiliation{Department of Physics, Boston College, 140 Commonwealth Ave, Chestnut Hill, Massachusetts 02467, USA}
%\author{Siming Wang}
 % \affiliation{Department of Physics and Center for Advanced Nanoscience, University of California, San Diego, La Jolla, California 92093, USA}
  % \affiliation{Materials Science and Engineering Program, University of California San Diego, La Jolla, California 92093, USA}

\author{Ilya Valmianski}
 \affiliation{Department of Physics and Center for Advanced Nanoscience, University of California, San Diego, La Jolla, California 92093, USA}

\author{J. G. Ramirez}
\thanks{Current address: Department of Physics, Universidad de los Andes, Bogot\'{a} 111711, Colombia}
  \affiliation{Department of Physics and Center for Advanced Nanoscience, University of California, San Diego, La Jolla, California 92093, USA} %\affiliation{Department of Physics and Institute of Nanotechnology, Bar Ilan University, Ramat Gan 52900, Israel}

  \author{Christian Urban}
  \affiliation{Department of Physics and Center for Advanced Nanoscience, University of California, San Diego, La Jolla, California 92093, USA}

\author{Ruidan Zhong}
 \affiliation{Condensed Matter Physics \& Materials Science Department Brookhaven National Laboratory Upton, NY  11973, USA}

\author{John Schneeloch}
 \affiliation{Condensed Matter Physics \& Materials Science Department Brookhaven National Laboratory Upton, NY  11973, USA}

\author{Genda Gu}
 \affiliation{Condensed Matter Physics \& Materials Science Department Brookhaven National Laboratory Upton, NY  11973, USA}

%\author{Ivan K. Schuller}
%\affiliation{Department of Physics and Center for Advanced Nanoscience, University of California, San Diego, La Jolla,
%California 92093, USA}

\author{Isaac Henslee}
 \affiliation{Montana Instruments, 151 Evergreen Dr., Bozeman, Montana, USA.}

\author{Kenneth S. Burch}
 \email{ks.burch@bc.edu}
\affiliation{Department of Physics, Boston College, 140 Commonwealth Ave, Chestnut Hill, Massachusetts 02467, USA}%

\date{\today}% It is always \today, today,

\begin{abstract}
Raman micro-spectroscopy is well suited for studying a variety of properties and has been applied to wide-ranging areas. Combined with \comment{Gavin, "tune-able" should be one word -- "tunable"} tuneable temperature, Raman spectra can offer even more insights into the properties of materials. However, previous designs of variable temperature Raman microscopes have made it extremely challenging to measure samples with low signal levels due to thermal and positional instability as well as low collection efficiencies. Thus, contemporary Raman microscope has found limited applicability to probing the subtle physics involved in phase transitions and hysteresis. This paper describes a new design of a closed-cycle, Raman microscope with full polarization rotation. High collection efficiency, thermal and mechanical stability are ensured by both deliberate optical, cryogenic, and mechanical design. Measurements on two samples, \BS{} and \VO{}, which are known as challenging due to low thermal conductivities, low signal levels and/or hysteretic effects, are measured with previously undemonstrated temperature resolution.

\end{abstract}

\pacs{Valid PACS appear here}% PACS, the Physics and Astronomy
\keywords{Raman microscope, Cryo Optic, Low vibration, High NA}%Use showkeys class option if keyword
\maketitle

%\begin{quotation}
%The ``lead paragraph'' is encapsulated with the \LaTeX\
%\verb+quotation+ environment and is formatted as a single paragraph before the first section heading.
%(The \verb+quotation+ environment reverts to its usual meaning after the first sectioning command.)
%Note that numbered references are allowed in the lead paragraph.
%
%The lead paragraph will only be found in an article being prepared for the journal \textit{Chaos}.
%\end{quotation}

\section{\label{sec:level1}Introduction}

Raman micro-spectroscopy is well suited for studying a variety of properties including chemical, magnetic, lattice, thermal, electronic, symmetry, and crystal orientation.\cite{PhysRevLett.110.107401,LOUDON:1964wy, Ferrari:2006p2362, weber_merlin,PhysRevB.82.064503,2011ApPhL98n1911Z,Beekman:2012ea,1980PhRvL..44.1604L,PhysRevB.82.064503,smith1971raman,cardona2000light,you2005quantum,menendez1984temperature,tian2016local} As such this technique has been applied to wide-ranging areas including chemistry,\cite{mccreery2005raman} physics,\cite{devereaux2007inelastic} materials science\cite{kumar2012raman}, and biology\cite{tu1982raman}. Many interesting phenomena only emerge at low temperatures\cite{Raman_BSCCO_naturephysics,Raman_MTI,Raman_CDW,PhysRevLett.114.147201}, and as such it is often highly desirable to measure and/or image a sample below room temperature. In addition, the temperature dependence of Raman features often reveals new information such as the strength of phonon anharmonicity crucial for thermal properties.\cite{zhang2011raman}  For organic samples, low temperatures can immobilize the material in a near-native state, revealing much more detailed information about the samples and their interaction spectra.\cite{Micheal_ref3,holt1992freeze,lima2001raman,souza2002raman,saraiva2008temperature,kuball2002measurement} Thus, numerous insights can be gained by measuring the temperature dependence of the Raman response.  Typically, temperature control requires the use of cumbersome and expensive cryogenic liquids. For Raman micro-spectroscopy this can be extremely challenging, due to a number of factors including the rapidly rising cost of helium, the low Raman cross sections (typically 10$^{-8}$ to 10$^{-12}$)\cite{Raman_cross_section_graphite,Raman_cross_section_si}, the requirement of high spatial  and spectral resolution, as well as the need to use low laser power to prevent heating. This often creates a competing set of requirements, long integration times, the use of high numerical aperture (NA) objectives with low working distances, minimized use of helium, the need to place the sample in vacuum, and the need to keep the objective at a fixed location/temperature. To date, this has led to two different designs of low-temperature Raman microscopes. The first approach is to place the objective inside the cooling medium/vacuum, which enables high NA, but requires cryo-compatible objectives and leads to strong temperature dependence of the objective's performance as well as its relative alignment with the sample. The second approach employs an intermediate N.A., long working distance, glass compensated objective outside the cryostat. This results in higher mechanical stability of the objective, but at the cost of the spot size, polarization and especially the collection efficiency.\cite{Micheal_ref3} In addition, one also desires to make systems as automated as possible to reduce the operational errors and enable higher temperature resolution. Therefore, there has been an increasing demand to implement an automated system with high collection efficiency, thermal and mechanical stability.

In this article, we describe a new Raman microscope design, equipped with automated cryogenic temperature, laser power and polarization control as well as motorized imaging functions. Temperature changes in our system are based on an automated closed-cycle Cryostation, designed and manufactured by Montana Instruments Inc.  For Raman excitation and collection, a Cryo Optic module was employed, comprising a 100X, 0.9NA microscope objective, installed inside the Cryostation and kept at a constant temperature by a proportional-integral-derivative (PID) control loop. An agile temperature sample mount was designed and installed, ensuring fast thermal response (less than 5 minutes from 4 K to 350 K) as well as excellent Cryostation platform mechanical (5 nm) and thermal (mKs) stability. In addition to the small spot size and excellent collection efficiency this enables the improved collimation by the objective, which provides excellent spectral resolution, stability, and Rayleigh rejection with Notch filters (allowing signals down to 30 \WN{}). Moreover, nearly perfect polarization response can be measured at any in-plane angle using a Fresnel rhomb. These combined features produce much more reliable measurements with long integration times and continuous experiments lasting multiple weeks without the need for human intervention. This opens the door to widespread use of cryogenic Raman microscopy to probe nano-materials with low thermal conductivities and  very weak Raman responses.

\section{\label{sec:operation_principles}Operating Principles}
Raman scattering or the Raman effect is the inelastic scattering of a photon.  The energy difference between the incoming and outgoing photons corresponds to an excitation energy within the measured materials, enabling the use of a single wavelength light source to probe multiple excitations of a material. The basic components of a Raman microscope comprise a continuous wave laser, optical components to guide and focus the beam onto the sample, a laser filter, and a detector.  After illuminating a sample, the elastic Rayleigh as well as Raman scattered light is collected  by the same objective in the backscattering configuration.\cite{ferraro2003introductory}  Usually, the collected light must pass through a set of filters before entering a spectrometer because of the very small scattering cross-section of the Raman processes.\cite{lin1991handbook}

In Raman experiments, one often employs group theory to determine the symmetry of the mode measured. Specifically, the mode intensity is given by a Raman tensor ($\textbf{R}$) and the polarization of the incoming ($\vec{e_{i}}$) and outgoing ($\vec{e_{o}}$) light. Indeed, the crystallographic axes are the bases of $\textbf{R}$ and the intensity is given by:
\begin{equation}\label{Raman_intensity}
I_{R}=|\vec{e_{o}}\cdot\textbf{R}\cdot\vec{e_{i}}^{T}|
\end{equation}
Therefore by measuring the Raman response for various configurations of $\vec{e_{i,o}}$ as well as at different angles with respect to the sample, one can gain insights into the symmetry of the excitations and determine the crystallographic orientation.\cite{Beekman:2012ea}

\section{\label{sec:level1}Instrumentation}

\begin{figure*}
  \centering
  \includegraphics[width=\textwidth]{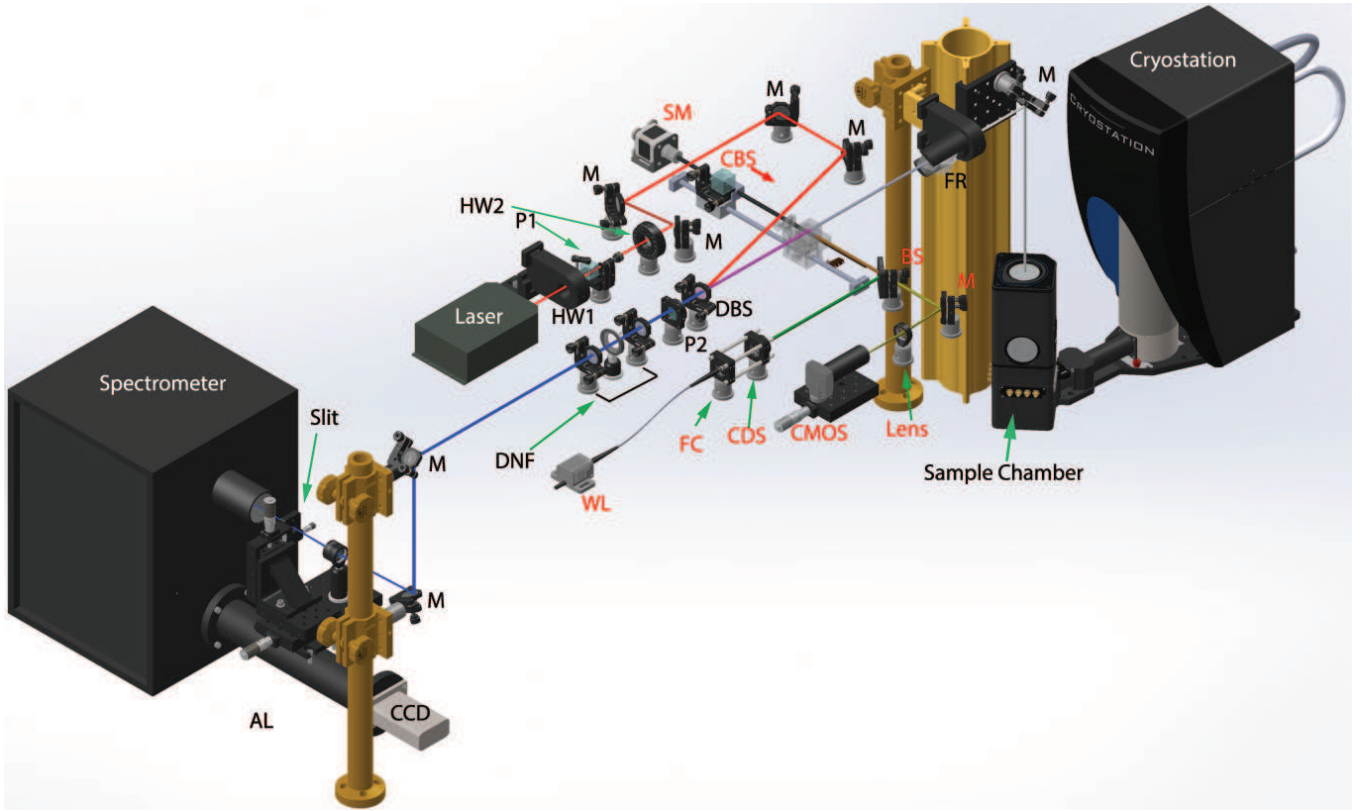}\\
  \caption{Model of our Raman cryo-microscope. The optical breadboard and the supporting frame are hidden for clarity. Laser and collected Raman signals are shown by the red and blue beams respectively. The shared beampath between the two are shown in purple. Components used to image the sample with white light  are denoted in red. The illumination and reflected imaging paths are shown by the green and gold beams. The shared beampath between the two are shown in brown and the one shared with the four is shown in light gray. Abbreviations refer to the following: \HW{} (HW), Mirror (M), polarizer (P), stepper motor (SM), cubic beam splitter (CBS), Fresnel rhomb(FR), diffractive beam splitter (DBS), beam splitter (BS), condenser (CDS) white light (WL), fiber coupler (FC), diffractive notch filter (DNF), achromatic lens (AL). }\label{fig:Raman_setup}
\end{figure*}

\subsection{Optical design}\label{optical:design}
We now describe in detail the overall design and layout of our Raman cryo-microscope system, which is shown in in FIG. \ref{fig:Raman_setup}). Our Raman setup starts with a laser source (the laser beam is shown by the red beams in FIG. \ref{fig:Raman_setup}). To achieve high spatial and spectral resolution, a Laser Quantum Torus 532 nm laser with a GHz-bandwidth was used for the excitation. A true zero-order half waveplate (HW1) mounted on a motorized universal rotator and a cubic polarizer (P1) were placed after the laser. Since the rotation of the half waveplate will induce a change in the polarization of the laser, the laser power after P1 can be effectively adjusted by rotating HW1, while maintaining the same polarization at the sample. In addition, the thinness of HW1 minimizes changes in the direction of the laser beampath upon changing the power. A second true zero-order half waveplate (HW2) was placed after P1, such that the measurement configuration could be switched between collinear (XX) and crossed (XY) polarizations. Following the optical path and several silver mirrors (M), the laser was directed to a diffractive 90/10 beamsplitter (DBS) from Ondax Inc, reflecting 90 \% of the excitation source, and rejecting 90\% of Rayleigh scattered light after exciting the sample. The laser is then guided towards a \MI{} Cryostation. Before entering the Cryostation, the laser passes a double \FR{} (Standa Inc., 14FR2-VIS-M27), which effectively acts as a broad-band half waveplate.

\begin{figure}
  \includegraphics[width=0.4\columnwidth]{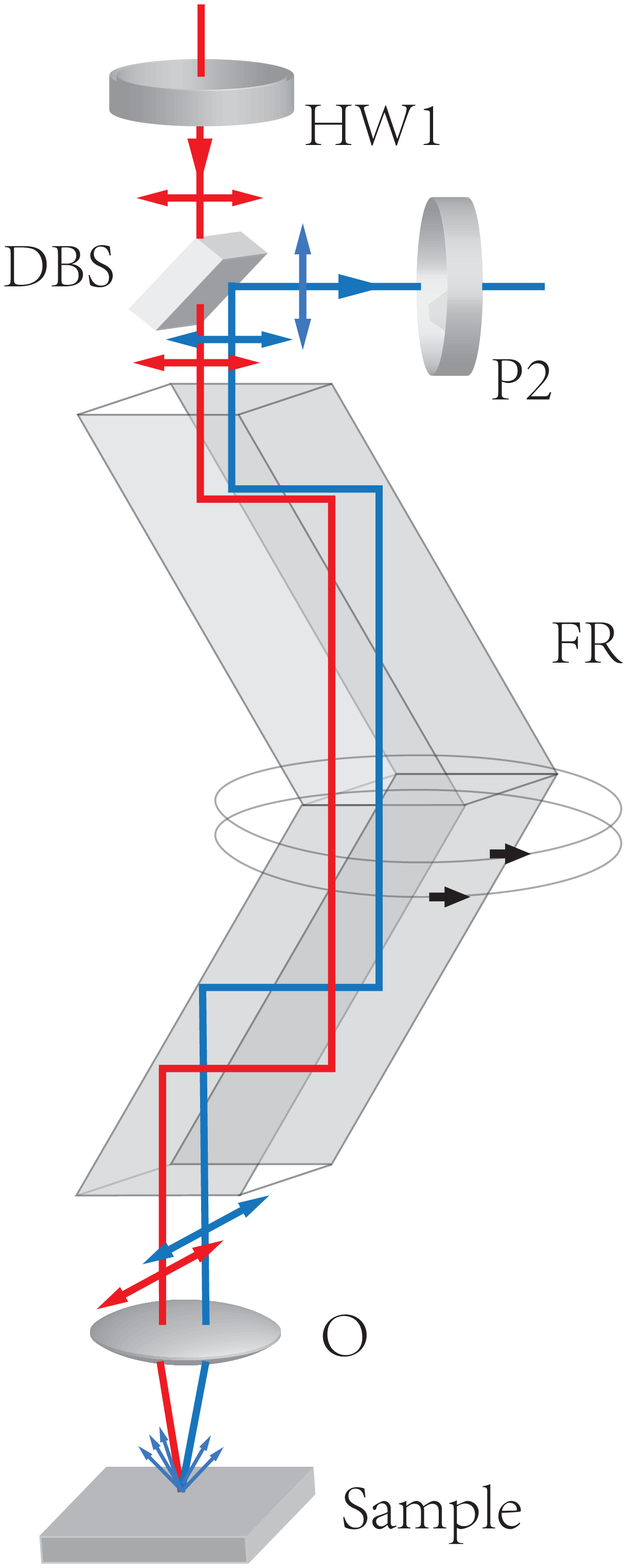}\\
  \caption{The light is total internal reflected four times and each time the relative phase shift $\pi/4$ between the s and p polarizations is added for a total $\pi$ phase. Thus, rotating the polarization of incoming laser(shown in red) can be achieved through the rotation of $\lambda/2$ Fresnel rhomb (FR).  The blue lines are both Raman scattered lights and Rayleigh  scattered lights.  Half waveplate (HW2), Analyzer (P2),  90/10 Beam splitter (DBS), Objective (O).\cite{Beekman:2012ea,kiefer2015universal,nestro1973circularly,kiefer2013determination} For clarity, the guiding mirrors between HW2 and DBS are omitted here. }\label{Rhomb_figure}
\end{figure}

After exciting the sample, Raman scattered light (shown in the blue beam in FIG. \ref{fig:Raman_setup}, see details in the caption of FIG. \ref{fig:Raman_setup} about the shared beampath) is collected by the objective inside the Cryostation and follows the incoming path. The Raman scattered light passes through the \FR{} and the analyzer (P2) after the 90/10  beamsplitter, as shown in FIG.\ref{Rhomb_figure}. As  mentioned above, the selection rules provide symmetry information about the excitation as well as the crystallographic axes. However, to achieve this without the \FR{}, one would have to rotate the sample, resulting in the loss of focus, lateral displacement of the beam with respect to the sample, and adding more complexity to the setup. Another method would be to simultaneously rotate components HW2 and P2 (Analyzer/polarizer). However, the diffraction grating in the spectrometer is most efficient for a fixed polarization (s-polarization, detailed discussion will be given in Sec. \ref{sec:charac}). Thus, rotating P2 would inadvertently affect the signal, even for an isotropic Raman tensor. The optimal result can be achieved by a double-loop search method. Basically, one can rotate HW2 by a small amount $\theta\approx5^o$, then rotate P2 while monitoring the signal level of the Rayleigh scattered lights to find the maximum counts, repeating until the counts are optimized. By introducing the \FR{}, one can fix P2 to optimize the efficiency of the spectrometer, manipulate HW2 to change from cross to co-polarized configuration and use the \FR{} to effectively rotate the sample's crystallographic axes about the fixed axes of the optical system.

To reject Rayleigh scattered light, and to reach a cut-off energy of 30 \WN,  two Ondax SureBlock volume holographic Bragg grating based diffractive notch filters (DNF) were placed after the analyzer. Using a notch filter as opposed to commonly used edge filters also allows both Stokes and anti-Stokes Raman signals to be recorded, which is a useful indicator to test local heating of the sample.\cite{jellison1983importance} These two filters plus DBS result in OD 8 attenuation to the Rayleigh scattered light. Finally, the Raman scattered light is directed to a Sine-drive spectrometer module equipped with an ultra-high resolution 2400 grooves/mm holographic grating. Prior to entering the spectrometer the light passes through an external mechanical slit mounted on a 3-axis translation stage. This allows for maximum resolution and collection efficiency by positioning the slit at the focal point of the first lens inside the spectrometer. To reduce chromatic and spherical aberration effects,  anti-reflection coated achromatic doublet lens (AL) with focal length of 50mm (Thorlabs, Inc. AC254-050-A-ML) was used to focus the light onto the slit. Ultimately, the diffracted light is detected by an Andor iDus back illumination spectroscopy CCD. The detector operates in sub-image bin mode and at the lowest rate ADC (analog-digital-conversion) channel to reduce the noise level.

To bring samples into focus during measurements, and for finding features and/or micron sized samples on a substrate, imaging the sample is necessary. To get more repeatable images, our Raman microscope is equipped with computer controlled  white light illumination and imaging capabilities (all parts denoted in red in FIG. \ref{fig:Raman_setup}). Illumination lights (shown by the green beam in FIG .\ref{fig:Raman_setup}) are delivered using a multi-mode fibre through a fiber coupler (FC) to a condenser (CDS) and then reflected by a beamsplitter (BS) towards a cubic beamsplitter (CBS).   A home-built motorized long range translation stage was used to move the CBS in/out of (shown by the transparent and opaque CBS respectively, in FIG .
\ref{fig:Raman_setup}) the beampath for imaging (Raman signal collection).  The CBS was fixed on a linear ball bearing carrier. A NEMA 17 stepper motor (SM) coupled with a 1/4-20 acme threaded shaft was used for the translation of the linear ball bearing carrier. The motion of the motor was controlled by a computer interfaced Arduino UNO microcontroller and a motor shield. Two microswitches connected to the UNO were used as stoppers for the carrier. When the CBS is moved into the beampath, the illumination lights share the same beampath as the laser and is directed onto samples.  After illumination, the reflected light followed the incoming path to the CBS, then is transmitted (shown by the gold beam in FIG. \ref{fig:Raman_setup}) through the BS and ultimately directed by a mirror (M) through a lens with 15 cm focal length to form an image on a Thorlabs CMOS camera (CMOS).

\begin{figure}
    \centering
 \includegraphics[width=\columnwidth]{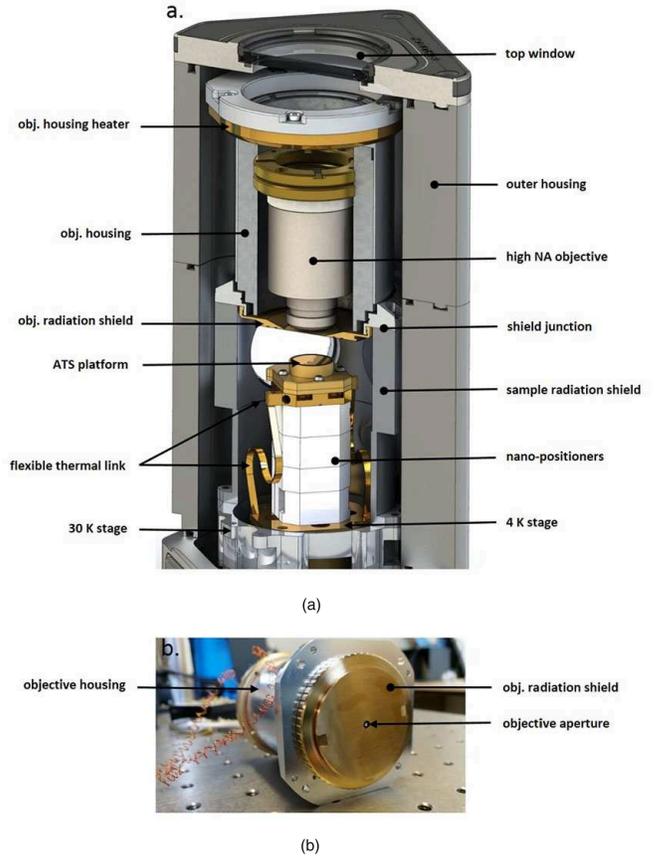}
 \caption{(a) Cross-section of the Cryo Optic module with Zeiss objective inside the Cryostation housing, highlighting critical components. (b) Bottom view of the objective housing showing a 1 mm beryllium copper aperture in detail from the point-of-view of the ATSM sample platform. The objective housing is thermally coupled to the 60 K sample radiation shield through a bolted interface.}\label{fig:cryochamber}
\end{figure}

%============================================================================================================
% FIGURE ATSM HEATLOAD MAP AND THERMAL CYCLES
%============================================================================================================
\begin{figure}
  \includegraphics[scale=0.37]{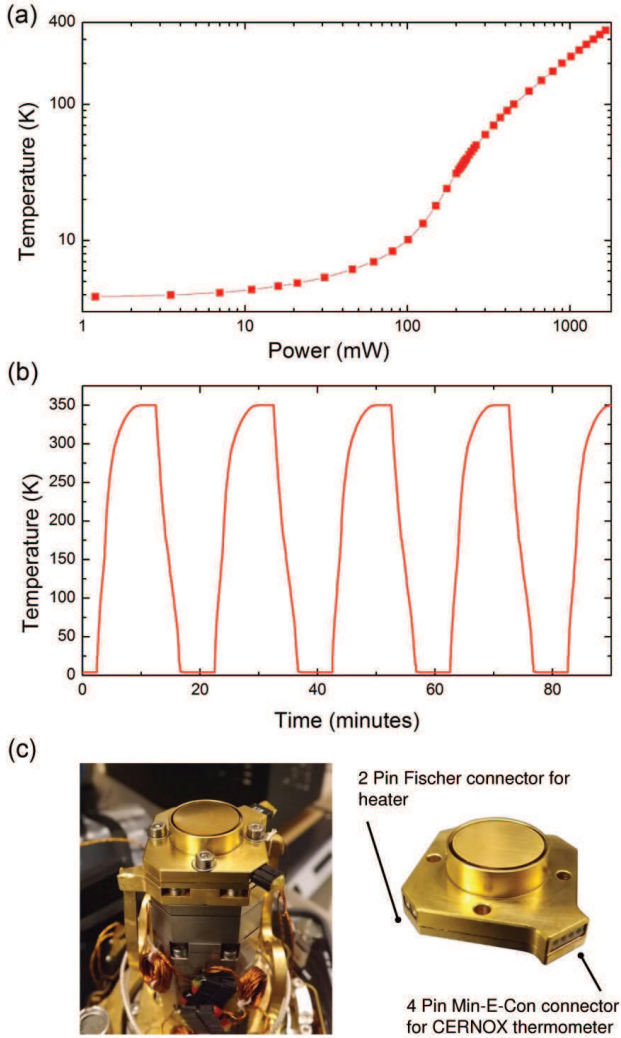}
  \caption{(a) Heatload map of the Agile Temperature Sample Mount, installed directly on the Cryostation platform. (b) 20 Minute thermal cycles of the ATSM between 4 - 350 K. (c) Pictures of the Agile Temperature Sample Mount illustrating the sample platform, the radiation shield and housing, and the thermometer and heater connectors. }\label{fig:heatloadcycle}
\end{figure}
%============================================================================================================

\subsection{Cryogenic and Mechanical Design}

%============================================================================================================
% FIGURE ATSM STABILITY
%============================================================================================================
\begin{figure}
  \includegraphics[scale=0.37]{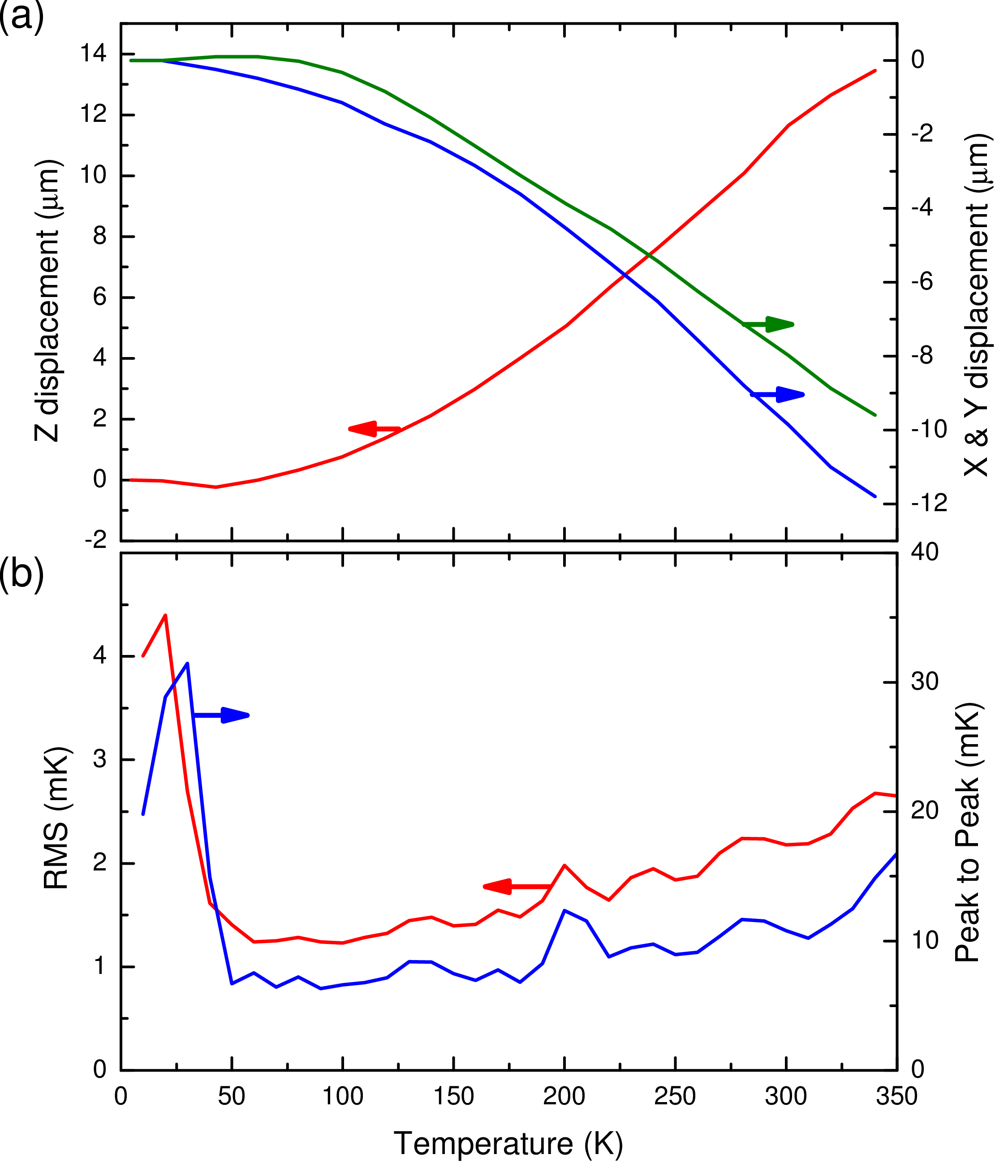}
  \caption{(a) Positional stability of the Agile Temperature Sample Mount between 4-350 K. (b) Thermal stability of the ATSM over 20 minutes per setpoint.}\label{fig:XYZstability}
\end{figure}
%============================================================================================================
\subsubsection{Cryostat with Cryo Optic}

To focus the excitation source on our sample, and collect (anti-) Stokes shifted radiation, we used a modified 100X, 0.90 NA Zeiss objective with a working distance of 310 $\mu$m, vented to operate in vacuum. The commercially available closed cycle Montana Instruments Cryostation is equipped with a Cryo Optic module that was developed in collaboration with the Burch group, and designed to mount the Zeiss objective. The Cryostation includes a low vibration ($\textless$5 nm peak-to-peak) cold platform ($\textless$3.5 K) for the sample, and a sample radiation shield (see FIG.  \ref{fig:cryochamber}). The radiation shield reduces radiative thermal loads on the sample, and serves as a thermal anchor for lagging wires that enter the sample space. Built-in Attocube 101-series xyz nanopositioners handle sample translation and focusing, providing 5 mm of translation range with sub-micron resolution. The Cryo Optic module mounts directly onto the radiation shield (as shown in FIG. \ref{fig:cryochamber}) so that the aperture separating the sample and the objective is cooled to 60 K. Moreover, the direct mechanical coupling of the objective and sample space inside the vacuum chamber ensure a stable focal plane with respect to the sample; an important requirement for long integration times, and difficult to achieve with an objective mounted outside of the vacuum chamber. This rigid mechanical connection is pivotal for the high NA objective, as the depth of focus is inversely proportional to the square of the NA\cite{hecht2002optics}. With the 532 nm excitation source, a spot size of $\approx 1~\mu m$ is achieved, as optically observed, and confirmed by the diameter of holes purposely burned into Bi$_2$Se$_3$ flakes. The combination of a small spot size, high NA, and high mechanical stability solves some of the key challenges in temperature dependent Raman investigations of novel samples as small crystal size, low heat capacity, and weak Raman signals require high lateral and axial resolution as well as low excitation power and thus long integration times.

A cross-section of the mechanical design of the Cryo Optic module is shown in FIG. \ref{fig:cryochamber}. Alignment of the objective with respect to the 50 $\mu$m thick beryllium copper aperture is achieved by attaching a reference surface to the bottom of the aperture plane and employing the previously described white light imaging capabilities of the system. The threaded copper rings to which the objective is mounted are rotated until the reference surface is in focus, and are subsequently locked in place. After installing the Cryo Optic module onto the radiation shield, the sample is brought into focus using the nanopositioners described above.

A key feature of the Cryo Optic module is the thermal isolation of the Zeiss objective and the 60 K aperture. This allows for a closed loop temperature controlled objective (kept at 310 $\pm$ 0.5 K) while the 60 K aperture limits the radiative heatload of the objective to a fraction of the Cryostation cooling power at 4 K. Moreover, with such careful temperature control, the objective and sample are impervious to typical environmental temperature fluctuations and the objective's optical performance is unaffected by the sample temperature. Hence, drift of the sample with respect to the objective is limited only by the thermal contraction and expansion of the Cryostation base platform, the nanopositioners, and the sample mount. To further reduce this drift, a sample stage was developed by Montana Instruments in collaboration with the Burch group that can change temperature nearly independent of the Cryostation platform and the nanopositioners, which is discussed in Sec. \label{sec:ATSM}.

\subsubsection{Agile Temperature Sample Mount}
\label{sec:ATSM}
To achieve high mechanical and thermal stability of the sample platform, an Agile Temperature Sample Mount (ATSM) was developed and integrated into the Raman microscope. FIG. \ref{fig:heatloadcycle}c shows two pictures of the ATSM, in which a 500 $\mu m$ thick copper sample platform is surrounded by a 4 K radiation shield, and is radially supported just below the platform by G-10 thermal stand-offs. This platform thickness and support geometry with rotational symmetry results in minimal drift in the focal plane of the microscope, and strongly enhances in-plane positional stability. For minimal error in sample temperature readings, a CERNOX CX-1050-HT is mounted on the bottom of the platform, thus separated from the sample by a 500 $\mu m$ layer of oxygen-free high thermal conductivity copper (OFHC). Moreover, the sample platform is equipped with closed loop controlled solid state heating elements (chip resistors in dead-bug configuration) for highly isotropic heat exchange and temperature control. This allows for precise temperature control over the sample platform, nearly independent of the Cryostation base platform, which maintains a temperature between 3-14 K over the full 4-350 K range of the ATSM. Hence, while cycling between temperatures, the thermal stability outside of the ATSM platform is only minimally affected, reducing thermal drifts within the system's radiation shield and housing over longer time-scales. As a result, a comparison of measurements taken before and after installing the ATSM revealed over an order of magnitude improvement in the time that is required to reach a stable temperature and position. This was determined both by temperature readings and producing a focused image between temperature changes. This is in part due to the agile temperature response of the stage (as described below), and in part due to the rapid thermal stability of the ATSM and optical train, as the temperature change and thermal expansions/contractions settle rapidly due to the minimal mass. FIG. \ref{fig:XYZstability}a illustrates the positional stability in the $xy$-plane and along the $z$-axis, where $z$ is parallel to the Poynting vector of the laser. These data were obtained by cooling down an NT-MDT Scanning Probe Microscope (SPM) calibration grating (model TGZ3) with a period of 3 $\mu m$. Using the closed loop positioners, a white light image of the grating was kept in focus and centered on the screen throughout the cooldown. All required {\it xyz} displacements as a function of temperature were recorded and plotted in FIG. \ref{fig:XYZstability}a.
%These data were obtained optically by thermally cycling an AFM calibration grating, while keeping the sample in focus and tracking features using closed loop positioners and recording their individual displacements.

Beyond overall stability, one may worry about vibrations caused by the closed-cycle system. This potential effect on the optical system was minimized by placing the optics on a separate platform from the Cryostation, though on the same optical table. The effect on the Raman signal was checked by collecting Raman spectra of a sample at room temperature with the Cryostation compressor turned on and with it turned off. No significant difference was observed. Furthermore, the overall design of the Cryostation and the ATSM minimizes the effects of vibration on the sample, as was measured using a Lion Precision CPL490 capacitive displacement sensor. With the ATSM mounted on top of the positioners stack, and with the Cryostation compressor running at high power, the room temperature in-plane ($x$ and $y$) vibrations did not exceed 60 nm, and are thus well below the diffraction limit of our 532 nm excitation source. While vibrations along the Poynting vector ($z$-axis) were not measured with the capacitive displacement sensor, they are unlikely to exceed the in-plane vibrations as a result of the positioner stage stack construction. White light imaging of the above mentioned SPM calibration grating (while the compressor was running) confirmed that mechanical vibrations along $z$ were below the axial resolution of our system as no defocussing was observed.  Vibrations of the ATSM were also measured in isolation of the positioners, while mounted directly to the Cryostation platform. In this arrangement, peak-to-peak vibrations did not exceed 5 nm, and an ATSM platform resonance frequency of 8.2 kHz was found.

The thermal stability of the ATSM platform is shown in FIG. \ref{fig:XYZstability}b, where each temperature point was recorded over a period of 20 minutes. We note that while the conventionally large sample platform mass passively aids thermal damping, the {\it low drift} geometrical constraints of the ATSM (and thus it's low mass) force it to strictly rely on a finely tuned PID loop for thermal stability, in addition to the thermal stability of the Cryostation. Indeed, with the low platform mass of the ATSM, and high thermal conductivity of OFHC copper below 50 K,\cite{NISTwallchart} we observed temperature changes $dT/dt$ in excess of 100 K/s at low temperatures, thus imposing stringent requirements on the frequency and I/O resolution of the PID control algorithm. A Lakeshore 335 temperature controller was used to meet these PID requirements. Our observed decline in thermal stability below 50 K coincides with a large jump in the thermal conductivity of copper, and is thus attributed to a sub-optimally tuned temperature control loop. Nevertheless, thermal stability over the full temperature range never exceeds 4.5 mK RMS (32 mK Peak-to-Peak), which easily meets the requirements for Raman microscopy, even at high temperature resolution, as shown in section \ref{sec:charac}.

It is important to note that while the ultimate goal of the ATSM is positional and thermal stability, its design also leads to a reduction in sample cooling power compared to the Cryostation base platform. This concern was addressed by optimizing the thermal standoff of the ATSM platform so that it can easily withstand typical radiative heatloads (0.1 mW) of the Raman excitation source by several orders of magnitude, while maintaining a low base temperature. FIG. \ref{fig:heatloadcycle}a shows a log-log heatload map of the ATSM over the full temperature range, emphasizing its low temperature cooling power. In addition to stability, the design of the ATSM also allows for highly agile temperature control, resulting in minimal time loss between temperature set-points. To illustrate this agility, FIG. \ref{fig:heatloadcycle}b shows a number of 20 minute thermal cycles between 4-350 K, stabilizing down to $<5$ mK RMS at both extremes of the cycle.

\section{Raman Characterization}
\label{sec:charac}
\begin{figure}
  \includegraphics[trim=2cm 6cm 2cm 6cm, width=\figuresize]{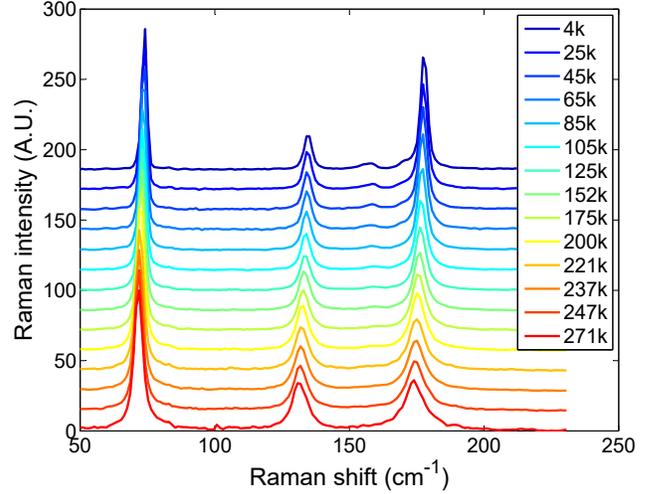}
  \caption{\label{Raman_temperature_BS}Temperature dependent Raman spectra of Bi$_{2}$Se$_{3}$ in XX configuration. Excellent signal to noise is observed at all temperatures, despite the low Raman response and thermal conductivity of Bi$_{2}$Se$_{3}$}
\end{figure}

\begin{figure}
  \centering
  \includegraphics[trim=3cm 9cm 3cm 9cm,width=\figuresize]{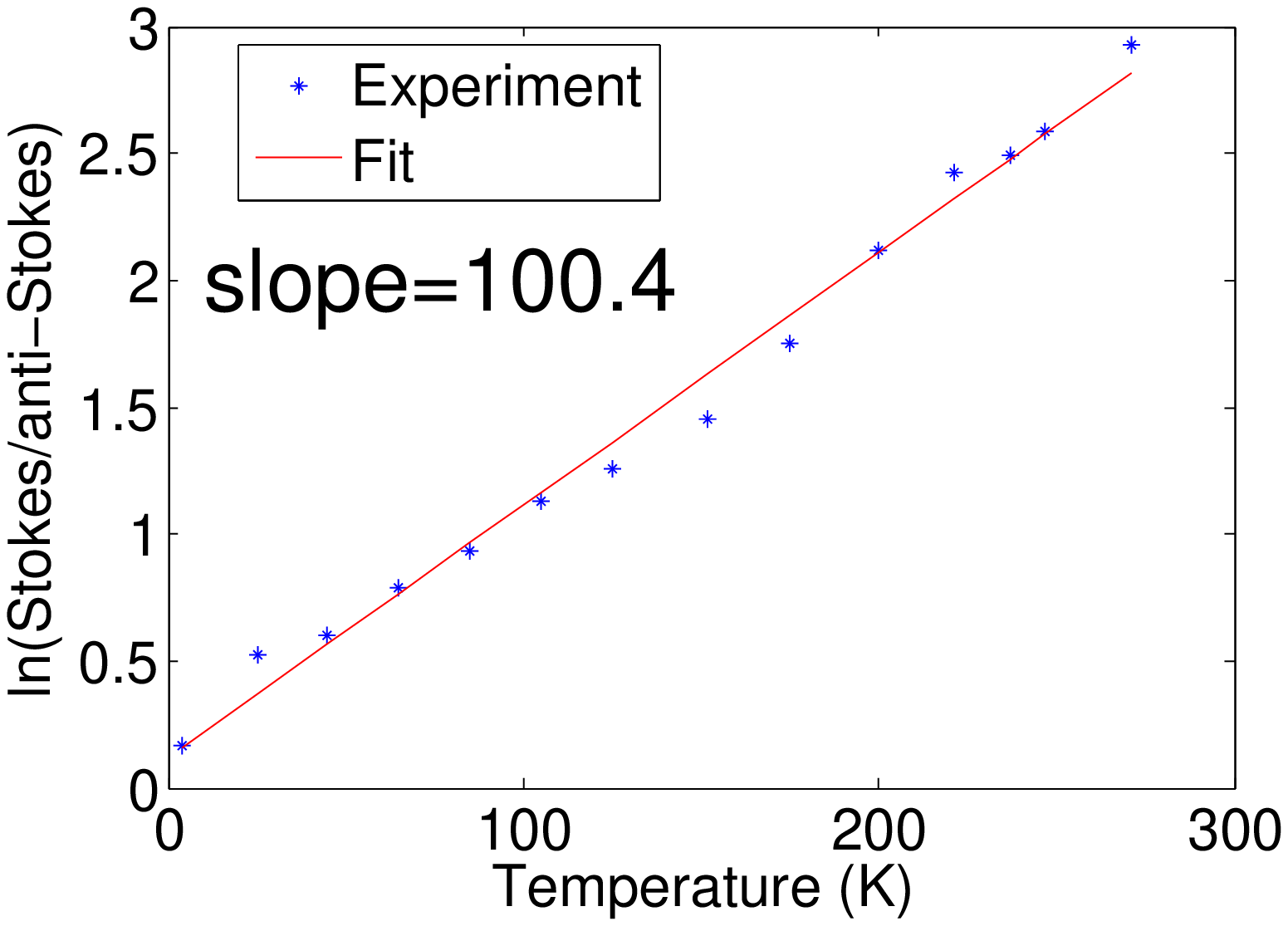}\\
  \caption{Temperature dependence of the inverse of logarithm ratio of Stokes and anti-Stokes of the first \BS{} phonon mode, indicating the absence of local heating.}\label{ratio_ANS_S}
\end{figure}

To demonstrate the performance of our setup, we measured two challenging samples, \BS{} and \VO{}. They are  of wide interest for applications in thermoelectric and memristive devices as well as their topological insulating (\BS{})\cite{FTIR_Be2Te3,laforge2010optical,hor2009p}  and strongly correlated behavior (\VO{})\cite{erekhinsky2013spin,de2014coercivity}. The first data set was obtained from single crystal \BS{} as it is of interest to a wide range of researchers such as those working in topological insulator, thermoelectric and nano materials.  Of particular interest is the potential for the temperature dependence of its phonons to unravel the origin of its low thermal conductivity.\cite{zhang2011raman} However the low thermal conductivity which makes it an excellent thermoelectric, also forces one to keep the laser power extremely low to avoid local heating. Thus, if the collection efficiency is not high enough, getting a sufficient signal to noise ratio is extremely challenging. Besides, the Raman response of \BS{} is well characterized, making it a good sample to test the capability of our system.\cite{:/content/aip/journal/apl/100/7/10.1063/1.3685465,PSSB:PSSB2220840226,zhang2011raman,Childres:2013hu,2011ApPhL98n1911Z,Shahil:2010fg}  The sample was freshly cleaved just before being placed into the sample chamber. The power of the laser used for the measurement was 40 $\mu$W to avoid the local laser heating and the laser spot size was $1~\mu m$ in diameter. The primary results are shown in FIG. \ref{Raman_temperature_BS}. Each spectra consists of an average of 5 acquisitions taken for 3 minutes each and is normalized by the height of the first phonon peak.
We can see there are three phonon peaks visible in the whole temperature range. They are located at 71.7 cm$^{-1}$, 131.5 cm$^{-1}$ and 174.0 cm$^{-1}$ at 271 K, in agreement with previous studies.\cite{PSSB:PSSB2220840226} To confirm the absence of local laser heating, we exploit the well known relationship between the ratio of the Stokes(S) and anti-Stokes(AS) intensity and the local temperature, given by: \cite{jellison1983importance}
\begin{equation}
R(S/AS)\propto{}exp(\frac{\hbar{}\omega_{0}}{k_{B}T})
\end{equation}
where $\omega_{0}$ is the phonon energy and $T$ the local temperature. Thus we expect a linear relationship between the sample temperature and the inverse of the log of this ratio. The absence of local heating is indeed demonstrated in FIG. \ref{ratio_ANS_S} where we plot the $log[R(S/AS)]^{-1}$ versus T for the lowest energy mode (71.7cm$^{-1}\approx 103~K$). This mode is chosen as it is the most sensitive to the temperature, nonetheless we find a good quantitative as well as qualitative agreement between the measured response and the prediction of no local heating.

\begin{figure}[!ht]
   \centering
    \includegraphics[width=\columnwidth]{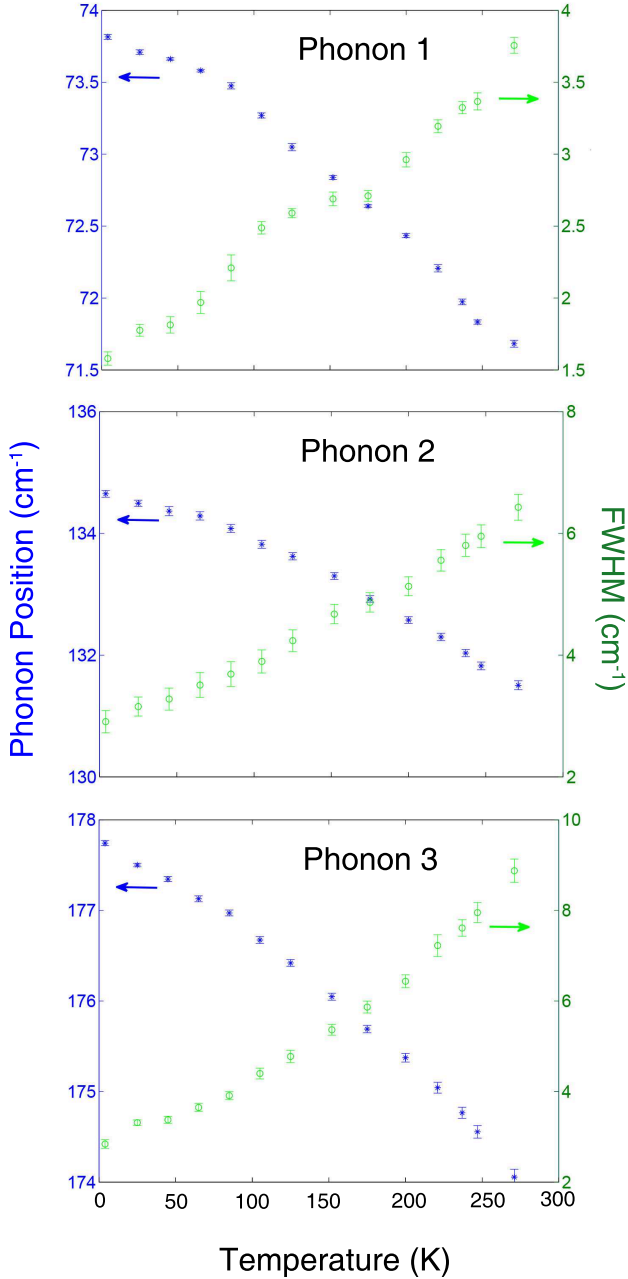}
    \caption{The temperature dependence of Bi$_{2}$Se$_{3}$ phonon shifts and linewidths. The errorbars in the plots are obtained from the fit (numerical regression algorithm).}
    \label{Phonon_position_width}
\end{figure}

\begin{figure}
  \centering
  \includegraphics[trim= 3cm 9cm 3cm 9cm,width=\figuresize]{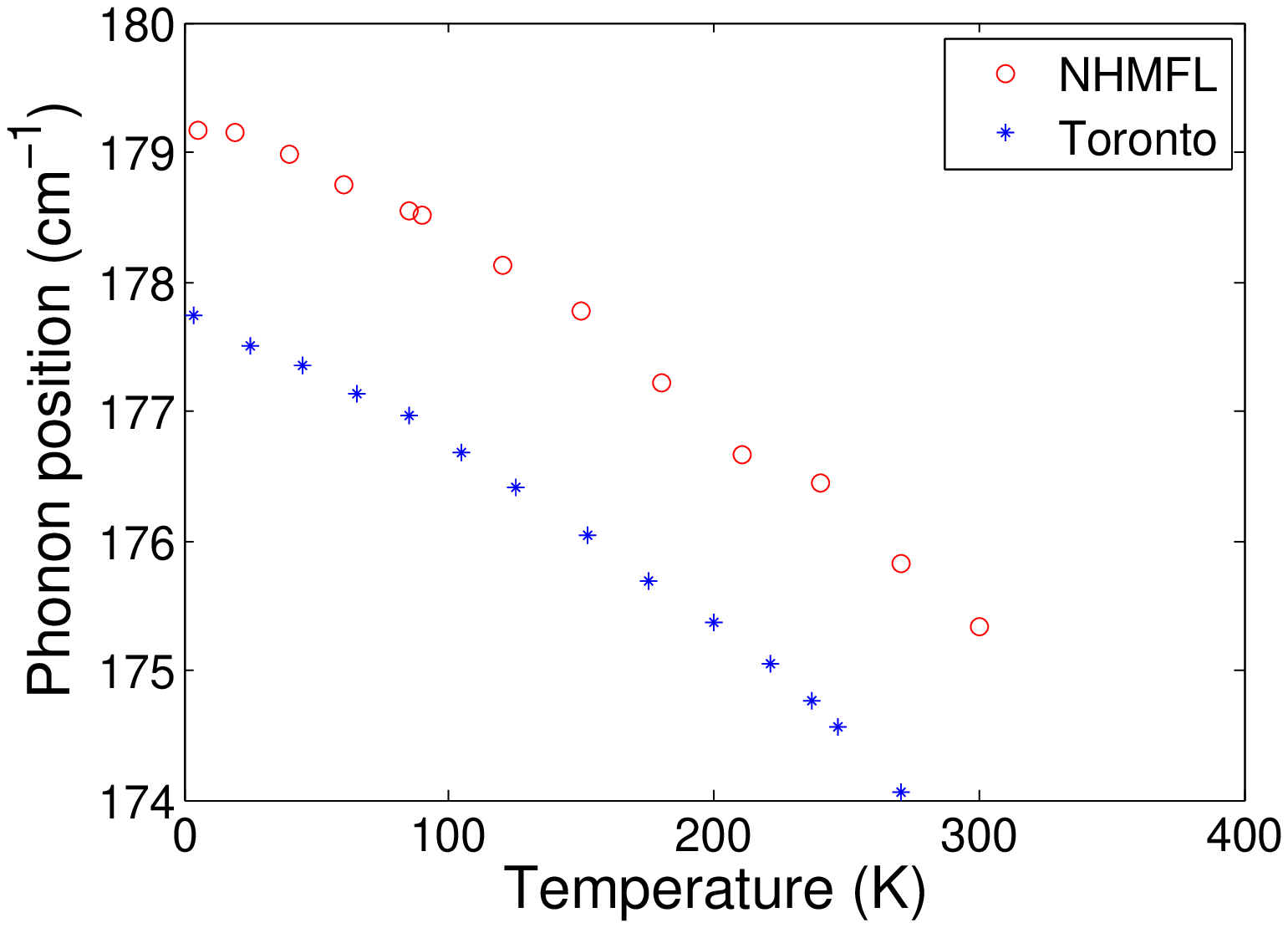}\\
  \caption{The temperature dependence of the third phonon of Bi$_{2}$Se$_{3}$. The data  in red are extracted from Ref.  \onlinecite{:/content/aip/journal/apl/100/7/10.1063/1.3685465}. The data in blue are identical to ``Toronto'' data and offset intentionally for comparison.}\label{comparison_BS}
\end{figure}

To further investigate the phonon temperature dependence, the phonon positions and linewidths were automatically extracted from raw spectra using a Matlab script programmed with a peak-detection function. The results are shown in FIG.\ref{Phonon_position_width}, where all three phonons are seen to harden and narrow as temperature decreases. This behavior matches well that found in a previous, independent study of a different Bi$_{2}$Se$_{3}$ crystal performed at the National High Magnetic Field Laboratory with an in-situ microscope objective based on fibre optics (see FIG.\ref{comparison_BS}).\cite{:/content/aip/journal/apl/100/7/10.1063/1.3685465} Comparing both data sets reveals a slight offset in phonon frequency, which is remedied by a temperature independent frequency shift, and can be accounted for by sample-to-sample variation and/or spectrometer calibration offset. Despite the use of free-space optics, random variation in our data set is much smaller than the results obtained by fibre optics which leads to a more smooth temperature dependence, confirming the thermal and mechanical stability of our setup and further improving the phonon analysis.

We now turn to our measurements of a \VO{} thin film. This material is of particular interest because under normal pressure, it  undergoes a first-order metal to insulator transition (MIT) and a structural phase transition.\cite{misochko1998optical,chen2012raman} Although the mechanism for the transition has been widely investigated using various experimental techniques such as transport, near-field infrared spectroscopy, ellipsometry, X-Ray, and various ultrafast methods,\cite{qazilbash2008electrodynamics,pfuner2005metal,liu2011photoinduced,abreu2015dynamic} the origin of the MIT is still controversial.\cite{ramirez2015effect} Variable temperature Raman spectroscopy could provide new insights into the temperature dependence of the volume fractions of each phase as well as the role of phonon-phonon interactions. However, due to the low signal level, nanometer sized domains and narrow temperature range in which the transition occurs, Raman measurements are very challenging and require very long integration times. Usually, for measurements like this, one would have to sacrifice the temperature resolution and signal to noise ratio for helium cost and thermal drifts. Furthermore, the transition is first order and occurs via percolation,\cite{wang2015avalanches} thus \VO{} exhibits extreme sensitivity to initial conditions as well as hysteresis, placing strict demands on thermal and positional stability. From this perspective, \VO{} is an ideal sample to test the overall capability of our setup from multiple angles, such as temperature resolution, collection efficiency, thermal and mechanical stability.

\begin{figure}
  \centering
  \includegraphics[width=\figuresize]{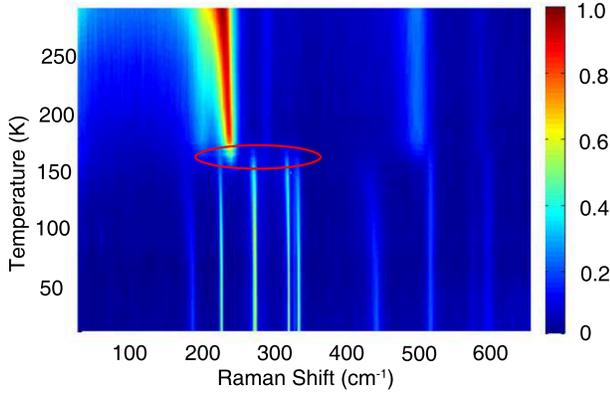}\\
  \caption{Temperature dependence of Raman intensity of \VO. The structural phase transition is easily observed as are the co-existence of the two phases.}\label{fig:Raman_V2O3}
\end{figure}

Temperature dependent Raman measurements were performed on a 200nm-thick \VO{} film grown on silicon.  We found excellent reproducibility between continuous thermal cycles measured at the same location on a \VO{} film over the course of one week. Specifically, each spectrum is an average of two to four 15 minute acquisitions with a laser power of 0.2 mW. The temperature resolution was chosen in the following way: below 140 K or above 180 K, the temperature steps were 15 K. In the two ranges of (from 140 K to 150 K and 168 K to 175 K), the temperature step was 2 K. Between 150 K and 168 K where the phase transition occurs, the temperature resolution was 1 K. The results are shown in FIG. \ref{fig:Raman_V2O3}.
The temperature dependent Raman spectra are consistent with a structural phase transition as has been reported  previously.\cite{ramirez2015effect} The metallic phase reveals both a broad asymmetric peak near 236 \WN{} and a low energy continuous scattered light by free carriers. The insulating phase, on the other hand, displays four strong modes located at 233.9 \WN, 279.4 \WN, 324.9 \WN and 340.4 \WN{} respectively. Interestingly, as expected for a first-order phase transition, we can also clearly see the co-existence of both insulating and metallic phases by the presence of both sets of phonons during the transition (circled region in FIG. \ref{fig:Raman_V2O3}). We also found hysteresis of the phase transition (to  be discussed in a future publication) by a significant difference in the onset temperature of the transition  when warming the sample.

\begin{figure}
\centering
\includegraphics[trim= 2cm 6cm 2cm 6cm, width=\figuresize]{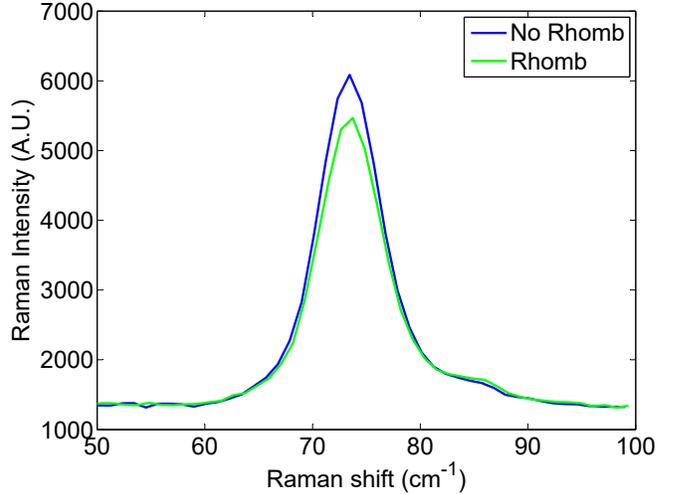}
\caption{Raman spectra of the first A$_{g}$ mode of \BS{} taken with (green) and without (blue) the \FR{}.}
\label{fig:rhomb_result}
\end{figure}

Lastly, we turn to another important characteristic in Raman spectroscopy; polarization dependence.  As mentioned in section \ref{optical:design}, another capability of our system is to vary the polarization of  incoming photons by introducing a \FR{}. This significantly increases the collection efficiency compared to the method of rotating both HW2 and P2. To confirm this, we have measured the the first phonon mode of \BS{} with and without the \FR{}, since the mode is of A$_{g}$ symmetry and has isotropic Raman response. The results are shown in FIG. \ref{fig:rhomb_result}. To get the loss ratio of Raman signal by introducing the \FR{}, we fit the phonon mode and took the ratio of the obtained amplitude in both cases. The calculation reveals the transmission efficiency is about $89\pm{}5\%$ which is much larger than the diffraction efficiency loss when the polarization changes from s-polarization (80\%) to p-polarization (less than 35\%) of the holographic grating.\cite{efficiecny_grating}

To demonstrate the polarization capabilities of the system, we performed a measurement on commercial grade single crystal silicon at room temperature. Single crystal silicon is of space group Fd$\bar{3}$m-227. The mode located at 520cm$^{-1}$ at room temperature is of T$_{2d}$ symmetry.\cite{Properties_of_crystalline_silicon} The Raman tensor of this mode is
\[R_{T_{2d}}= \left( \begin{array}{ccc}
0 & d & 0 \\
d & 0 & 0 \\
0 & 0 & 0 \end{array} \right)\]

\begin{figure}
 \centering
  \includegraphics[trim=2cm 6cm 2cm 6cm, width=\figuresize]{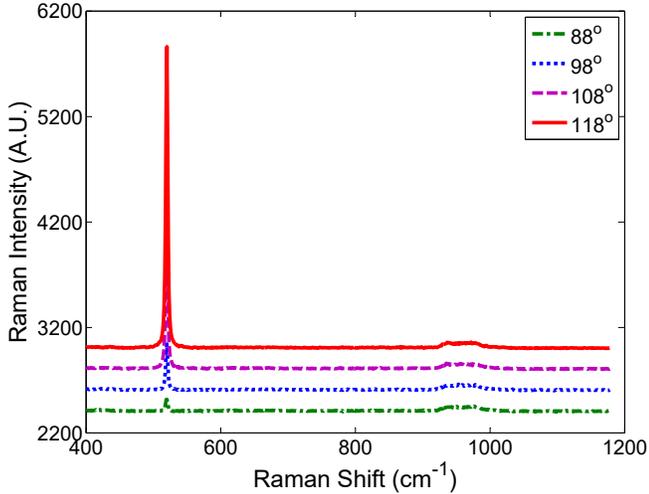}\\
  \caption{Polarization angle dependence of Raman spectra of silicon in XX configuration. For clarity, only four spectra are shown. Legend: Polarization angle.}\label{silicon_raman_spectra}
\end{figure}

\begin{figure}
 \centering
  \includegraphics[width=\figuresize]{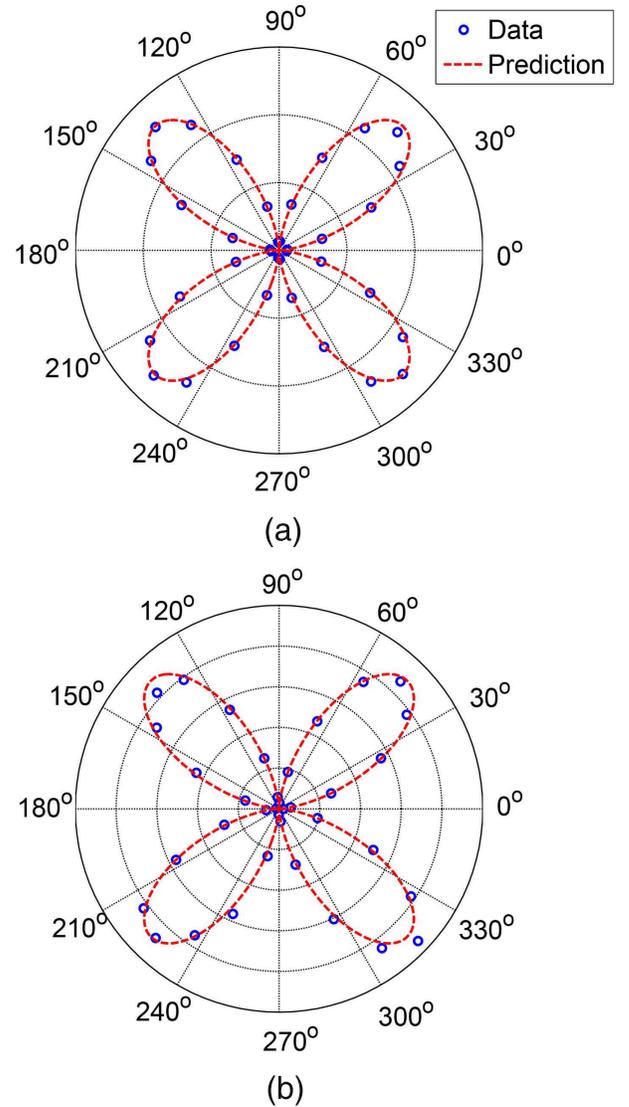}
  \caption{Polarization angle dependence of the relative Raman Intensity of silicon in XX (top) and XY (bottom) configuration.}\label{silicon_polarization_dependence}
\end{figure}

Using equation \ref{Raman_intensity}, one can show the Raman intensity should vary as $I_{R}=|sin(4\phi)|^{2}$, where $\phi$ is the polarization angle relative to the crystal axis in both XX and XY configuration. The measurement results are obtained at room temperature. Each spectra was taken under laser power 0.1 mW with an acquisition time of 20 s. The raw spectra are shown in FIG. \ref{silicon_raman_spectra}. From the figure, we can see two main features in each spectra, a one-phonon mode located at 520 cm$^{-1}$ and a weak two-phonon mode located from 900 cm$^{-1}$ to 1000 cm$^{-1}$. The one phonon mode varies with the change of polarization angle and the two-phonon modes barely change, consistent with the literature.\cite{lu2005polarization} To confirm the relation between Raman intensity and polarization angle derived above,  the area under the one phonon mode peak was extracted by the same fitting technique mentioned previously and then was plotted versus the polarization angle in FIG. \ref{silicon_polarization_dependence}. From the figure, we can see the agreement between the measurements and the theoretical prediction is excellent.

\section{Conclusion}
\label{sec:conclusion}
We have presented a new design of a temperature-dependent, closed-cycle, automated, micro-Raman spectroscopy system with high collection efficiency, and excellent temperature and mechanical stability. It enables long measurement times with very high temperature, spatial and spectral resolution. The closed-cycle nature also provides ease of use and minimal operation cost.
Given the automation potential of our setup, functions such as dynamical tracking of laser power level, auto-focusing and unattended operation will be developed, in the future. We have demonstrated a highly efficient, versatile automated micro-Raman cryogenic measurement platform.

\section{Acknowledgments}
We would like to thank Kerry Neal at Montana Instruments Inc. for technical help and insightful discussions.
Work at the University of Toronto was supported by NSERC, CFI, and ORF. K.S.B. acknowledges support from the National Science Foundation, (grant DMR-1410846)
Work performed at Brookhaven was funded through Contract No. DE-SC00112704.
\VO{} thin films fabrication and characterization in Ivan K. Schuller's lab at UCSD was supported by  the AFOSR Grant No. FA9550-12-1-0381. I.K.S. thanks the U.S. Department of Defense for support from a National Security Science and Engineering Faculty Fellowship (NSSEFF).

\section{References}
%\label{sec:ref}
%\bibliography{My_Collection}

\end{document}